# Far-field subwavelength imaging from a *single* broadband antenna in combined with strongly disordered medium

**Lianlin Li,[1,*] Fang Li,[2] and Tie Jun Cui[3]**

[1]*School of EECS, Peking University, Beijing, 100871, China*
[2]*Institute of Electronics, Chinese Academy of Sciences, Beijing, 100080, China*
[3]*State Key Laboratory of Millimeter Waves, Southeast University, Nanjing, 210096, China*
[*]*lianlin.li@pku.edu.cn*

The far-field subwavlength imaging is a challenging issue. In this letter we demonstrate numerically that the far-field subwavelength imaging of weakly scattering objects can be obtained by processing the data acquired by a *single* antenna, which benefits from the use of the strongly disordered medium. A mathematical model has been proposed for solving such problem based on the idea of sparse reconstruction. Moreover, this study leads to an important conclusion that the strongly disordered medium can serves as an efficient apparatus for the single-antenna compressive measurement, which shifts the complexity of devising compressive sensing (CS) hardware from the design, fabrication and electronic control. The proposed method and associated results can find applications in several imaging disciplines, such as optics, THz, RF or ultrasound imaging.

Over past years, a family of strategies has gained researchers' intensive attentions, which explores the multiply scattering of wavefields in a strongly disordered medium (or scattering medium) to enhance the quality of focusing and imaging in various applications, such as acoustics, microwave, optics, etc [1-5]. For these schemes, an array of detectors is mandatorily required to obtain an imaging with high spatial resolution, for instance, the well-known time-reversal mirror (TRM in short). Recently, in the context of broadband time reversal, Pierrat et al investigated theoretically the feasibility of subwavelength focusing in an open disordered medium with a *single* antenna [1]. This letter considers a more general imaging problem: to achieve the subwavelength imaging of sparse specimen from a single antenna. We demonstrate numerically that with the aid of the strongly disordered medium, the subwavelength imaging of sparse reflectors can be obtained by processing the data acquired by a *single* antenna at far field. This study leads to a concept of single-antenna subwavelength imaging, which allows us to beat the Rayleigh limit for imaging from the far-field. Compared to existing technologies for subwavelength imaging, our method employs a single antenna for data acquisition, and thus comes with several potential advantages. First, it is helpful for real-time imaging application. Second, it makes no use of near-field measurements, mechanical scanning, or phase shifters.

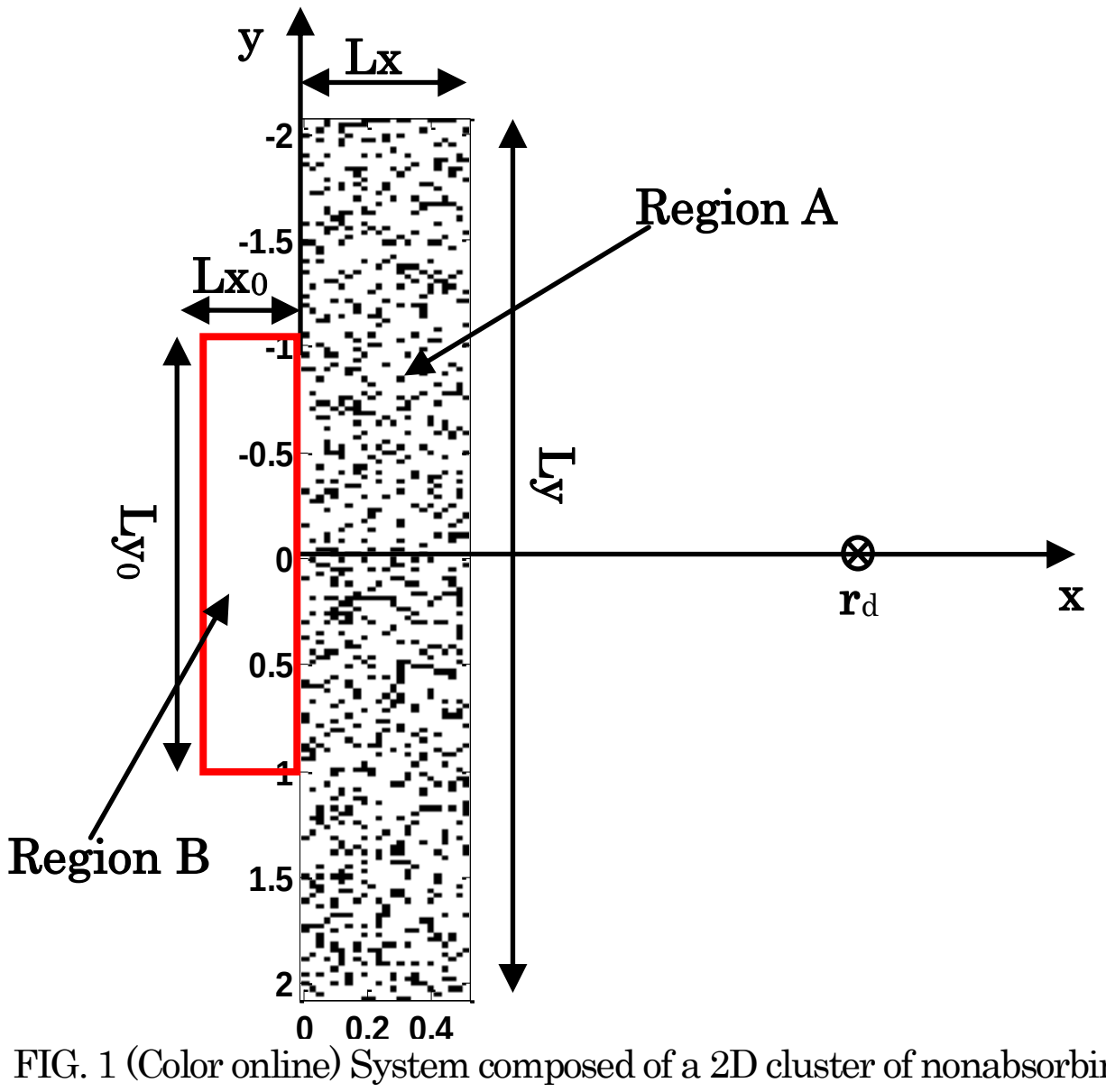

FIG. 1 (Color online) System composed of a 2D cluster of nonabsorbing cylindrical scatters randomly placed inside the rectangle region **A** with size of Lx by Ly. The probed broadband sources lie inside the rectangle region **B** with size of $Lx_0$ by $Ly_0$. Detection with a single point sensor is located at $\mathbf{r}_d$ in the far field. The radius of cylindrical scatters is 22nm, relative permittivity $\varepsilon_r$ =8, and volume fraction 15%.

It is worth highlighting that the system, consisting of the disordered medium in combined with single antenna, can act as an apparatus for real-time compressive measurements. Over past years, the theory of compressive sensing (CS in short) introduces an elegant paradigm shift in signal acquisition, which provides a remarkable reduction in the number of measurements without loss of reconstruction fidelity [6, 7]. CS demonstrates that the reduced significantly number of measurements, accomplished by acquiring an incomplete set of pseudo–random projections, is sufficient to recover a sparse or compressible signal under test by employing a tractable nonlinear reconstruction algorithm. Several imagers based on the spirit of CS were constructed, such as the single-pixels camera [8], the random lens [9], the meta-imager [10], and so on. On practical grounds, these CS implementations require the sequential generation of a large number of random patterns, and thus remain inefficient. This study introduces a novel CS technique based on a disordered medium that enables subwavelength imaging of general broadband sources from a single antenna. Contrarily to aforementioned CS techniques, our single-antenna sensing method makes no use of a sequence of random patterns, thus strongly reducing acquisition time.

We consider a two-dimensional (2D) sensing problem (or so-called inverse source problem [e.g., 11, 12]) to demonstrate the working principle of proposed methodology. As sketched in Fig. 1, a 2D cluster of nonabsorbing cylindrical scatters distributed randomly inside a rectangle region A of Ly=4.0μm by Lx= 0.5μm. The radius of cylindrical scatters is 22nm, and a minimum distance 5nm is forced between scatters to avoid overlapping. For numerical simulation, the operational wavelength varies from 538nm to 838nm with a separation of 5nm. Both these cylindrical scatters and objects under test are aligned infinitely along z-axis, and the illuminated sources are polarized along the scatters (TE mode), such that the electromagnetic problem is scalar. These objects under test lie inside the rectangle region B with size of $Ly_0$=2.0μm by $Lx_0$=0.2μm. Note that $Lx_0$=0.2μm is chosen so as to preserve adequate near-filed interaction between probed objects and scatters. A single point antenna, which is used to emit broadband illumination and receive echoes from objects under test, lies in the far field, i.e., at position $\mathbf{r}_d$= (6.8, 0) μm. In our simulations, the region B is uniformly divided with N=124 squares of 0.688nm by 0.688 nm. To avoid introducing too many notations, the symbol of $\mathbf{r}_s$ denotes the location of the *s*th (s=1, 2,…,N) pixel in the rest of this letter.

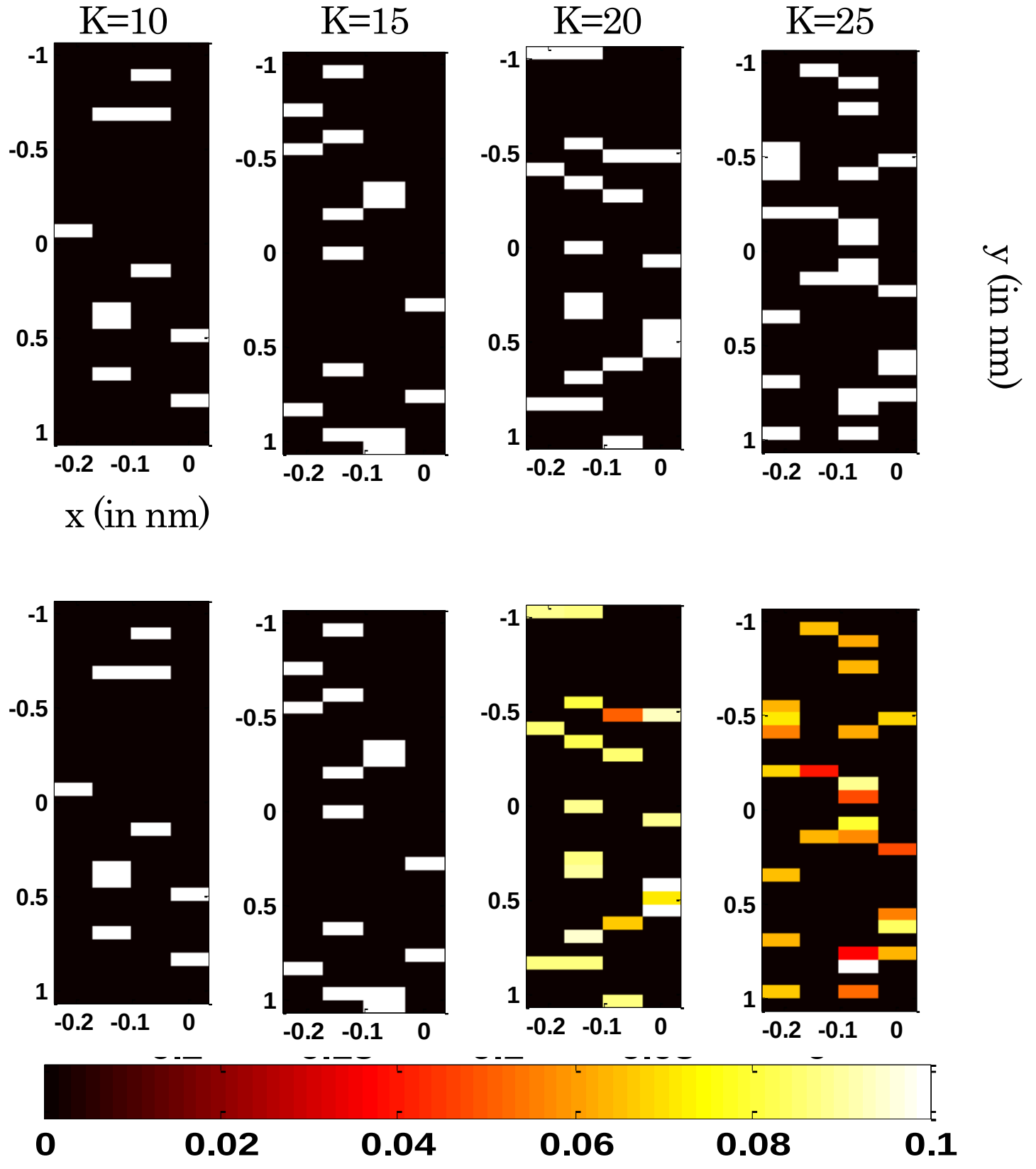

FIG. 2. (Color online) The normalized intensities reconstructed by proposed methodology for varying number of non-zero objects K=10, 15, 20, and 25. The first row is for the ground truth, while the second row for the reconstruction. The x-axis denotes the location along x-direction in nm, while the y-axis is for y-direction in nm.

Considering weakly scattering objects, the scattering field, $y(\omega)$, from objects can be represented via the Born approximation as [11, 12]

$$y(\omega) = k_0^2 \sum_s G^2(\mathbf{r}_d, \mathbf{r}_s; \omega) O(\mathbf{r}_s) \qquad (1)$$

where ω denotes the angular frequency, $O(\mathbf{r}_s)$ denotes the reflectivity of the *s*th pixel in the imaged scene, the square accounts for mono-static effects. In Eq.(4), $G(\mathbf{r}_d, \mathbf{r}_s; \omega)$ is the Green's function for the disordered medium system, see details in Appendix. Then, our goal is to reconstruct $O(\mathbf{r}_s)$ by solving Eq.(1). Generally, Eq. (1) is ill-posed since the independent measurements available are seriously inadequate relative to the number of unknowns. In some practical cases, the vector formed by $O(\mathbf{r}_s)(s = 1,2,\dots)$ is sparse, which implies that only a small fraction of it is nonzero or significant. Inspired by CS, Eq. (1) is regularized through the $l_1$-norm of $O(\mathbf{r}_s)$, then one derives following sparsity-constraint optimization problem:

$$\min_{p(\theta;\omega)}\{\int[y(\omega) - k_0^2\sum_s G(\mathbf{r}_d, \mathbf{r}_s; \omega)O(\mathbf{r}_s)]^2 d\omega + \gamma\sum_s|O(\mathbf{r}_s)|\} \qquad (2)$$

where γ is a factor used to balance the data fidelity and regularization term. The well-developed iteratively reweighted algorithm is performed to solve Eq. (2) [14, 15] in this letter.

Figs. 2 display reconstructions of $O(\mathbf{r}_s)$ by solving Eq. (5) for a varying number of non-zero reflectivity of $O(\mathbf{r}_s)$, where the ground truths are provided in the top row, and the reconstructions are shown in the bottom row. In our simulations, the values for non-zero weakly reflectors $O(\mathbf{r}_s)$ are set to be 0.1, and the SNR (signal-to-noise) is 30dB. Moreover, the minimum separation between two adjacent dipoles is set to be 0.1λ of the minimum resolved distance. From this set of figures one can see that the sparse weak objects can be easily distinguished in the subwavelength spatial scale. It is also noted that when the probed objects are not sparsely distributed, their locations still can be clearly indentified although their values of reflectivity deviate from truths. It is worth highlighting that with a single broadband antenna, the objects distributed in the two dimensional space can be accurately retrieved in the subwavelength scale. It can also be expected that the sparse objects in the three dimensional (3D) space can also be faithfully recovered in the subwavelenth scale from a single broadband antenna.

Now we examine the effect on reconstruction quality from operational wavelength range centered at 688nm, and to investigate the capability of proposed methodology for locating the number of weak objects. Fig. 3 shows the MSEs of reconstructions as a function of the operational wavelength range R, and the number of probed objects K, which is obtained and averaged over 50 Monte-Carlo (MC) independent trials. In our MC experiments, the locations of K probed objects under test are randomly selected from the whole N pixels. The normalized MSE is defined as

$$\mathrm{MSE} = \frac{\sqrt{\sum_s[\hat{O}(\mathbf{r}_s) - O(\mathbf{r}_s)]^2}}{\sqrt{\sum_s[O(\mathbf{r}_s)]^2}} \qquad (3)$$

where $\hat{O}(\mathbf{r}_s)$ is the reconstructed reflectivity, and $O(\mathbf{r}_s)$ corresponds to truths. From above results, one reasonable conclusion can be immediately deduced that the higher the value of K is, the worse the successful probability is.

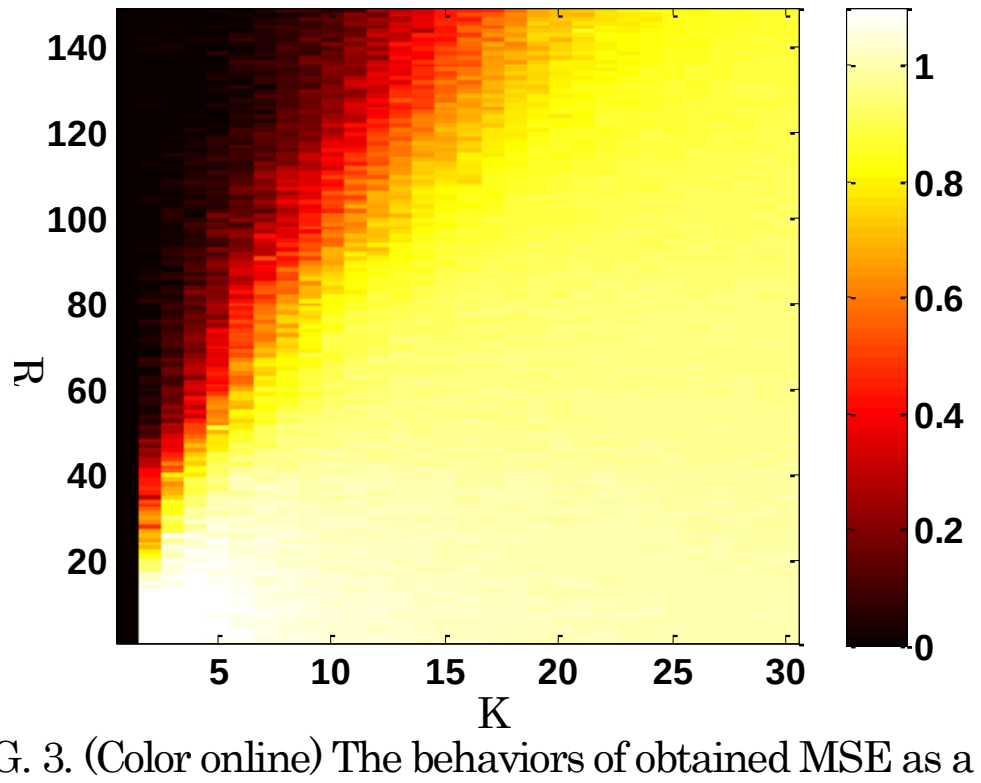

FIG. 3. (Color online) The behaviors of obtained MSE as a function of the operational wavelength range centered at 688nm, R, and the number of non-zero objects, K. Note that the range of working wavelength is 688nm+[-R/2, R/2]nm. (The Matlab code for reproducing Figs. 3 can be freely achieved by sending a request email to lianlin.li@pku.edu.cn)

To look insights into above conclusions furthermore, we re-express Eq. (1) in the form of linear equations, namely,

$$\begin{bmatrix} y(\omega_1) \\ \vdots \\ y(\omega_t) \\ \vdots \\ y(\omega_T) \end{bmatrix} = \begin{bmatrix} G^2(\mathbf{r}_d, \mathbf{r}_{s,1}; \omega_1) & \cdots & G^2(\mathbf{r}_d, \mathbf{r}_{s,n}; \omega_1) & \cdots & G^2(\mathbf{r}_d, \mathbf{r}_{s,N}; \omega_1) \\ \vdots & \vdots & \vdots & \vdots & \vdots \\ G^2(\mathbf{r}_d, \mathbf{r}_{s,1}; \omega_t) & \cdots & G^2(\mathbf{r}_d, \mathbf{r}_{s,n}; \omega_t) & \cdots & G^2(\mathbf{r}_d, \mathbf{r}_{s,N}; \omega_t) \\ \vdots & \vdots & \vdots & \vdots & \vdots \\ G^2(\mathbf{r}_d, \mathbf{r}_{s,1}; \omega_T) & \cdots & G^2(\mathbf{r}_d, \mathbf{r}_{s,n}; \omega_T) & \cdots & G^2(\mathbf{r}_d, \mathbf{r}_{s,N}; \omega_T) \end{bmatrix} \begin{bmatrix} O_1 \\ \vdots \\ O_n \\ \vdots \\ O_N \end{bmatrix} \qquad (4)$$

where T and N denotes the total number of discrete frequencies and image pixels for numerical computations. Among all the features that were proposed to characterize a matrix as a good candidate for CS, the coherence plays a special role because of its extreme easy-implementation [15]. For this consideration, Fig. 4 compares the coherences for the measurement matrix in Eq. (4) with that associated with the one of the Gaussian random matrix intensively discussed in the CS literatures. The similar behavior confirms that the disorder medium can be a good candidate in a CS setup for subwavelength imaging of sparse objects.

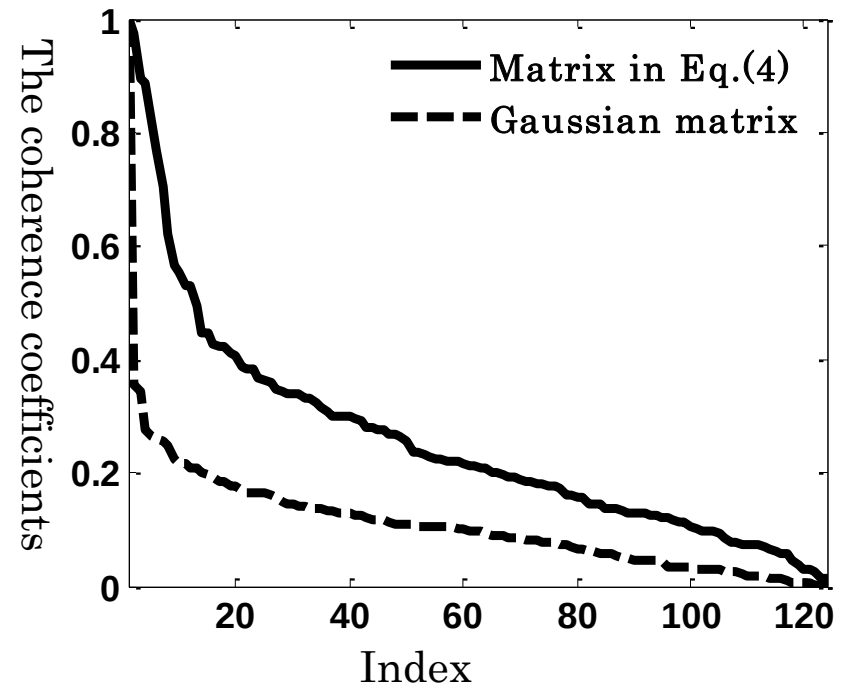

FIG.4 (Color online) The coherence coefficients of mapping matrix in Eq. (4) and the randomly generated Gaussian matrix, which is sorted in the descend manner.

Above results demonstrate that the sparse objects can be reconstructed from a single broadband antenna in combined with a disordered medium. Now we turn to discuss a more general problem: what's the performance for the object which itself is not sparse, but has sparse representation in a transformed domain. For simplicity, the Haar wavelet is considered hereto after. Several reconstruction samples are reported in Figs. 5, where the operational wavelength varies from 538nm to 838nm with a separation of 5nm. As can be seen, the objects being sparse in the Haar domain also can be almost perfectly reconstructed from a single broadband antenna.

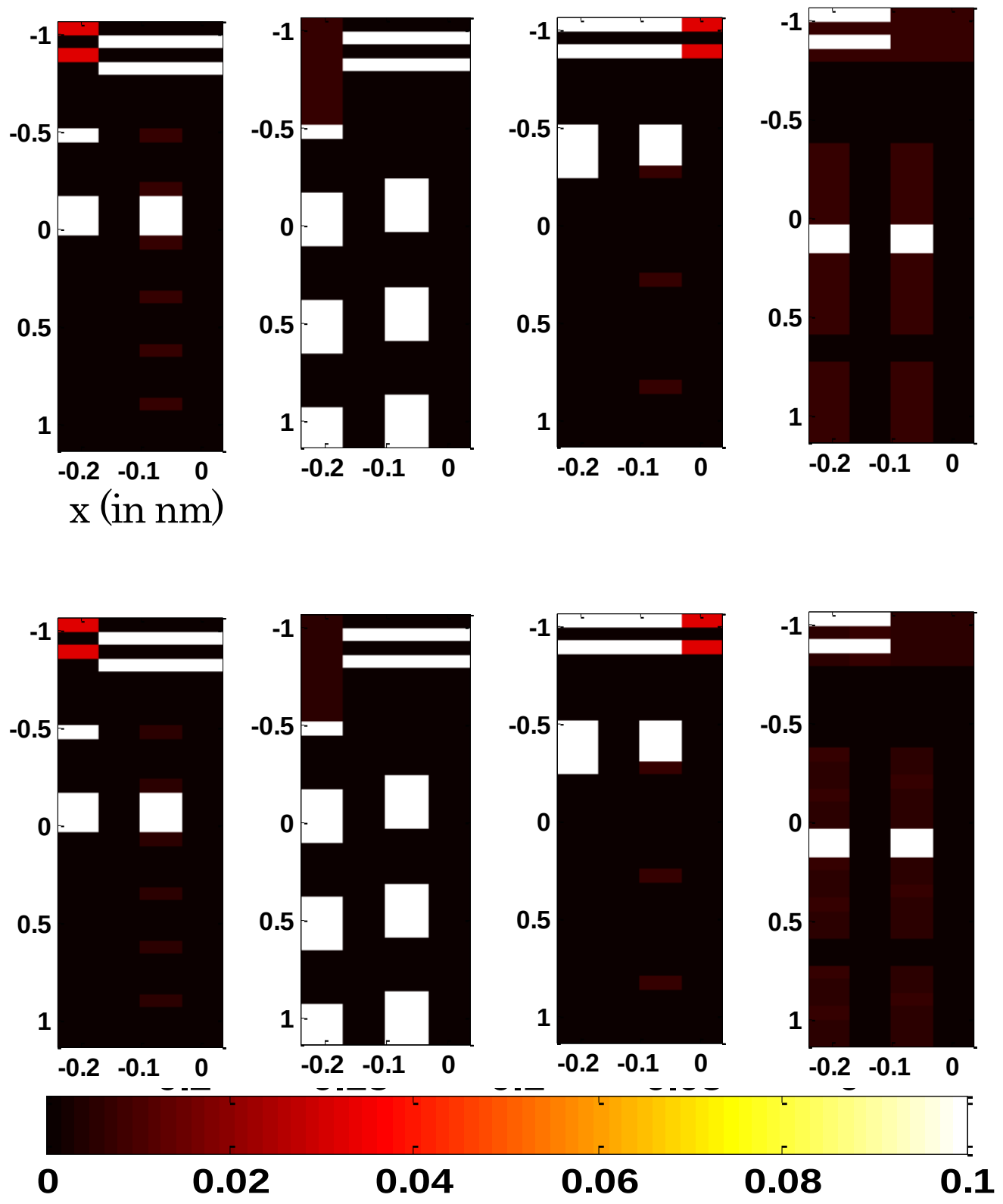

FIG. 5. (Color online) Four reconstruction samples of objects being sparse in the Haar-wavelet domain. The first row is for the ground truth, while the second row for the reconstruction. The x-axis denotes the location along x-direction in nm, while the y-axis is for y-direction in nm.

In summary, we develop a novel scheme of the far-field subwavelength imaging from a *single* antenna in combined with a strongly disordered media. We show heuristically that a strongly disordered medium can serves as an efficient apparatus for compressive measurement, which shifts the complexity of devising CS hardware from the design, fabrication and electronic control to a simple and single calibration procedure. Furthermore, unlike most current CS hardware, this system gives access to full compressive measurements by one single antenna, drastically speeding up acquisition. Moreover, this imaging device makes no use of any conventional lens, and thus such implementation can find applications in other disciplines, such as, THz, RF or ultrasound imaging.

**References**

1. R. Pierrat, C. Vandenbem, M. Fink and R. Carminati, Physical Review A, 87, 041801(R), 2013
2. I. M. Vellekoop and A. P. Mosk, Optics Letters, 32.16, 2309-2311, 2007
3. I. M. Vellekoop, A. Lagendijk, and A. P. Mosk, Nature Photonics, 4.5, 320-322, 2010
4. L. Li and F. Li, Applied Physical Research, 4, 30-41, 2010
5. A. Liutkus, D. Martina, S. Popoff, G. Chardon, O. Katz, G. Lerosey, S. Gigan, L. Daudet, and I. Carron, arXiv. 1309, Sep., 2013
6. E. J. Candes, J. Romberg, and T. Tao, IEEE Trans. Information Theory, 52, 480-509, 2006
7. D. Donoho, IEEE Trans. Information Theory, 52, 1289-1306, 2006
8. M. F. Duarte, M. A. Davenport, D. Tarkhar, J. N. Laska, S. Ting, K. F. Kelly, and R. G. Baraniuk, IEEE Signal Processing Magazine, 25, 83-91, 2008
9. R. Fergus, A. Torralba, and W. T. Freeman, Random lens imaging, Tech. Rep. Massachusetts Institute of Technology, 2006
10. J. Hunt, T. Driscoll, A. Mrozack, G. Lipworth, M. Reynolds, D. Brady, and D. R. Smith, Science, 339, 137-313, 2013
11. W. C. Chew, Waves and fields in inhomogeneous media, IEEE Press, 1995
12. M. Born, and E. Wolf, Principles of optics, Ed.7, Cambridge Press, 1999
13. P. C. Chaumet, A. Sentenac, and A. Rahmani, Physical Review E, 70, 036606, 2004
14. L. Li and B. Jafarpour, Inverse Problem, 26, 105016, 2010
15. M. Elad, Sparse and redundant representations: from theory to applications in signal and image processing, Springer Press, 2010

Appendix:

First, we derive the Green's function $\mathrm{G}(\mathbf{r}_\mathrm{d}, \mathbf{r}_\mathrm{s}; \omega) = \mathrm{E}(\mathbf{r}_\mathrm{d}, \omega; \mathbf{r}_\mathrm{s})$for the whole disordered medium system, which relates the electric field at the position of antenna $\mathbf{r}_\mathrm{d}$ to a given point source at $\mathbf{r}_\mathrm{s} \in$ Region B. To proceed, we employ the coupled-dipole method for solving following coupled electrical integral equations [13], i.e.,

$$\mathrm{E}(\mathbf{r}_\mathrm{d}, \omega; \mathbf{r}_\mathrm{s}) = \mathrm{E}^0(\mathbf{r}_\mathrm{d}, \omega; \mathbf{r}_\mathrm{s}) + k_0^2 \chi \sum_{\mathrm{j}=1}^{\mathrm{J}} \mathrm{E}(\mathbf{r}_\mathrm{j}, \omega; \mathbf{r}_\mathrm{s}) \int_{\mathrm{V}_\mathrm{j}} \mathrm{G}_0(\mathbf{r}_\mathrm{d}, \mathbf{r}'; \omega) \mathrm{d}\mathbf{r}' \qquad \text{(A.1)}$$

and

$$\mathrm{E}(\mathbf{r}_\mathrm{i}, \omega; \mathbf{r}_\mathrm{s}) = \mathrm{E}^0(\mathbf{r}_\mathrm{i}, \omega; \mathbf{r}_\mathrm{s}) + k_0^2 \chi \sum_{\mathrm{j}=1}^{\mathrm{J}} \mathrm{E}(\mathbf{r}_\mathrm{j}, \omega; \mathbf{r}_\mathrm{s}) \int_{\mathrm{V}_\mathrm{j}} \mathrm{G}_0(\mathbf{r}_\mathrm{i}, \mathbf{r}'; \omega) \mathrm{d}\mathbf{r}' \qquad \text{(A.2)}$$

$$\mathrm{i}=1,2,\ldots,\mathrm{J}$$

where $\chi = \varepsilon_\mathrm{r} - 1$, $J$ is the total number of cylindrical scatters in region A, and $\mathrm{G}_0(\mathbf{r}, \mathbf{r}'; \omega)$=j/4$\mathrm{H}_0^{(1)}(k_0|\mathbf{r} - \mathbf{r}'|)$ is the free space Green's function with $\mathrm{H}_0^{(1)}$the first-kind and zero-order Hankel function. Herein we assume that the wavefiled is uniform over one scatter, which is a good approximation if the scatter is smaller than the operational wavelength $\lambda$. $k_0$is the wavenumber dictated by $k_0 = 2\pi/\lambda$. Let

$$\mathrm{E}^0(\mathbf{r}_\mathrm{i}, \omega; \mathbf{r}_\mathrm{s}) = \mu_0 \omega^2 \mathrm{G}_0(\mathbf{r}_\mathrm{i}, \mathbf{r}_\mathrm{s}; \omega) \qquad \text{(A.3)}$$

be the electrical field associated with an electromagnetic wave impinging on the $i$th scatter from a source located at $\mathbf{r}_\mathrm{s}$. Figure A map the normalized amplitude (top row) and phase (bottom row) of the Green's function $\mathrm{G}(\mathbf{r}_\mathrm{d}, \mathbf{r}_\mathrm{s}; \omega)$ associated with the one configuration as shown in Fig. A, where the x-axis denotes the index of image pixels, while the y-axis is for the operational wavelength in nm.

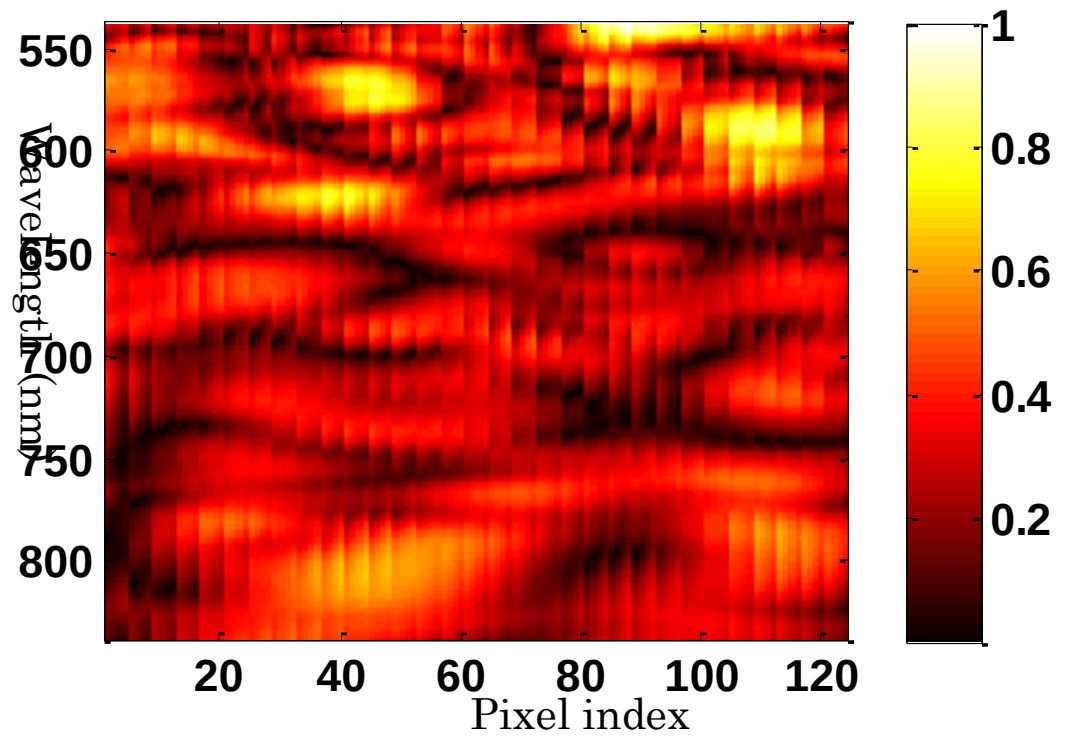

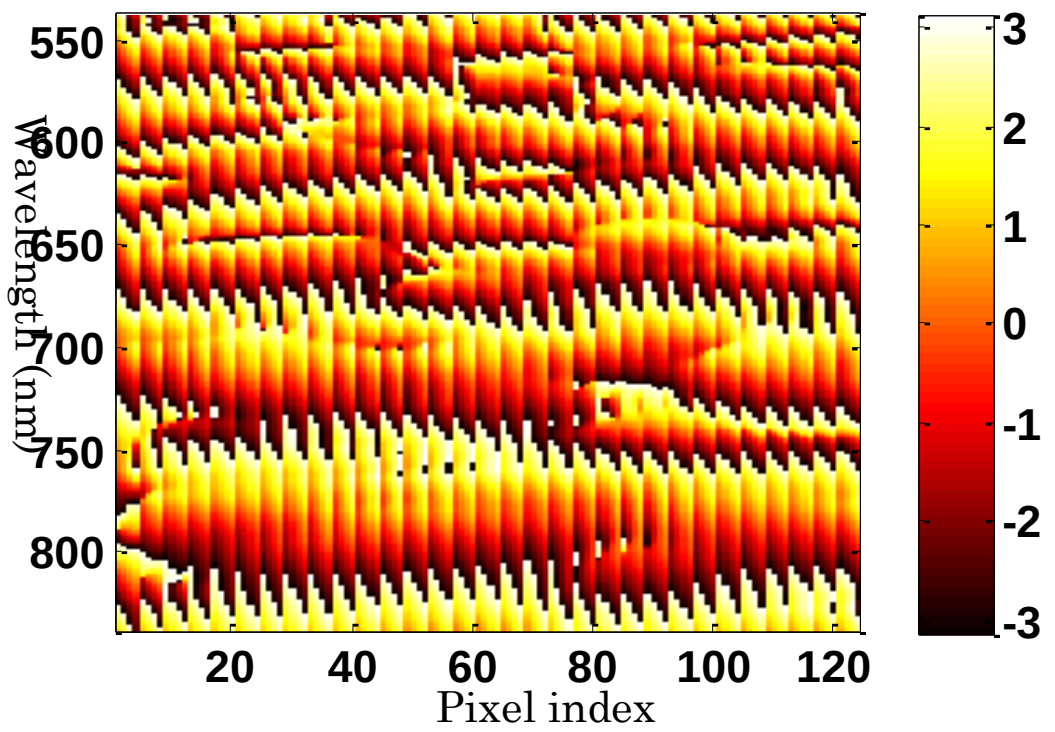

FIG. A. (Color online) The normalized amplitude (top row) and phase (bottom row) of the Green's function $G(\mathbf{r}_d, \mathbf{r}_s; \omega)$ associated with the one configuration as shown in Fig. 1. In this set of figures, the x-axis denotes the index of image pixels, while the y-axis is the operation wavelength in nm.